\documentclass[11pt,english]{article}
\usepackage[a4paper]{geometry}
\usepackage{amsmath}
\usepackage{amssymb}
\usepackage{graphicx}
\usepackage{setspace}
\usepackage{wasysym}
\makeatletter

\newcommand{\lyxaddress}[1]{
\par {\raggedright #1
\vspace{1.4em}
\noindent\par}
}

\date{23 June 2015}

\makeatother

\usepackage{babel}
\begin{document}

\title{Friction Causing Unpredictability}

\author{Joshua Oldham and Stefan Weigert}

\maketitle

\lyxaddress{\begin{center}
Department of Mathematics, University of York\\
York YO10 5DD, United Kingdom
\par\end{center}}
\begin{abstract}
The periodic motion of a classical point particle in a one-dimensional
double-well potential acquires a surprising degree of complexity if
friction is added. Finite uncertainty in the initial state can make
it impossible to predict in which of the two wells the particle will
finally settle. For two models of friction, we exhibit the structure
of the basins of attraction in phase space which causes the final-state
sensitivity. Adding friction to an integrable system with more than
one stable equilibrium emerges as a possible ``route to chaos''
whenever initial conditions can be specified with finite accuracy
only. 
\end{abstract}

\section{Introduction}

The difficulty to reliably predict the behaviour of a classical systems
is usually related to the existence of fractal structures in the mathematical
model describing the system. Conservative non-integrable systems such
as three interacting planetary bodies \cite{poincare99} and chaotic
dissipative systems such as Lorenz's model of the atmosphere \cite{lorenz63}
provide two well-known cases in point. Consider adding a third body
to the integrable system of two planetary bodies which interact through
gravitation. The KAM theorem \cite{arnol'd89} describes how the original
foliation of the system's phase-space into tori is being replaced
gradually by a highly intricate mixed phase space. Finite balls of
initial conditions will no longer contain trajectories on tori only
but also others which separate at an exponential rate. The non-linearity
present in Lorenz's model gives rise to a strange attractor \cite{ruelle+71}.
Its properties dominate the long-term evolution of the system since
trajectories with neighboring initial conditions are likely to visit
rather different regions of phase space at comparable later times.
This phenomenon has been called \emph{final state sensitivity} \cite{grebogi+83}.

An actual macroscopic physical system, however, cannot exhibit fractal
structures in a strict sense: the system would need to match its mathematical
description on \emph{arbitrary} length scales \cite{mandlbrot77}
but classical models break down on the molecular or atomic level.
Experimentally observed structures may be highly intricate over many
-- but not all -- orders of magnitude. Nevertheless, finite intricacies
are sufficient to make reliable long-term predictions impossible when
combined with initial states which are known only approximately. 

Adding friction to conservative non-integrable model systems washes
out frac\-tal struc\-tures on the finest scales but remnants may continue
to exist. This has been shown for a spherical pendulum with three
stable equilibrium positions, in the presence of gravity \cite{motter+13}.
When friction is added, basins of attraction with highly intricate
boundaries emerge. If the initial state is not known exactly the fine
structure of the basin boundaries -- in spite of not being fractal
-- already prevents any reliable prediction of the equilibrium position
near which the pendulum will come to rest.

The purpose of this paper is to identify a different mechanism which
also results in an unpredictable final state if initial conditions
are known only with finite accuracy. Starting from an \emph{integrable
}system\emph{ }with multiple stable equilibria, we will show that
the addition of friction may \emph{create }basins of attraction with
intricate boundaries, leading to a situation which resembles the one
of the pendulum just described. The main difference is that this mechanism
does not rely on pre-existing fractal structures: instead, incorporating
friction \emph{causes} intricate structures to emerge in an originally
integrable system. 

Sec. \ref{sec:Final-state-sensitivity} of this paper provides an initial, qualitative explanation of why adding friction to an integrable system with two or more stable equilibria may cause final-state sensitivity. Then, in Sec. \ref{sec:Model-systems}, we investigate the structure of basins of attraction generated by two different types of friction when acting on a particle moving in a piece-wise constant double-well potential. Finally, we summarize and discuss our results in Sec. \ref{sec:Summary}.

\section{Final-state sensitivity in a double-well potential\label{sec:Final-state-sensitivity}}

\begin{figure}[h]
\begin{centering}
\includegraphics[bb=150bp 0bp 1200bp 500bp,scale=0.5]
{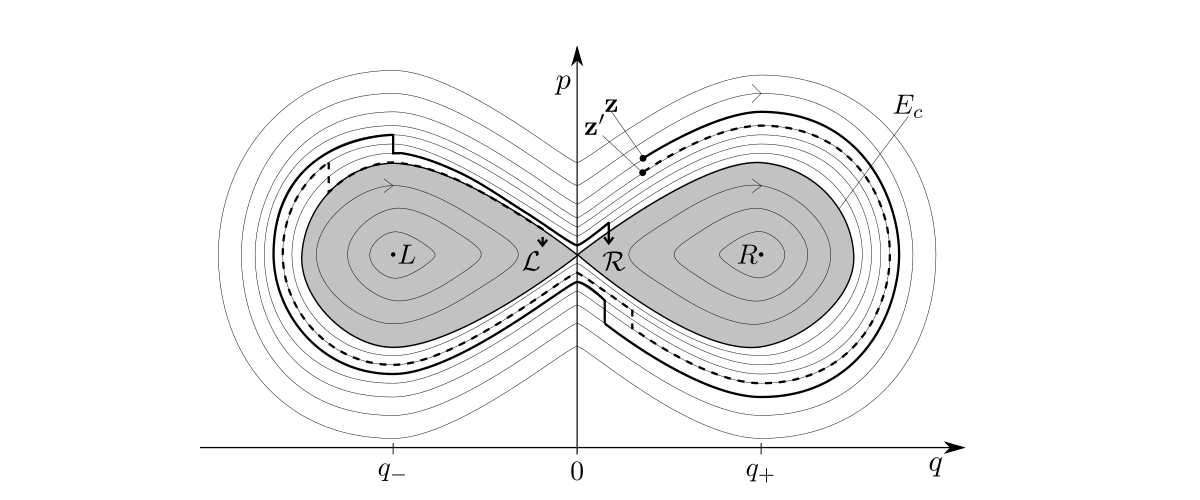} \protect\caption{Phase-space structure of a double-well potential $W(q)$ with minima
$L$ and $R$, located within wells ${\cal L}$ and ${\cal R}$ (shaded
areas), respectively; a particle with neighbouring initial conditions
$\mathbf{z},\mathbf{z}^{\prime}$ (full and dashed lines, respectively)
may end up in different wells when subjected to periodic, dissipative
``kicks'' which reduce its momentum and thus its energy. \label{fig: double well flow (general)} }
\end{centering}
\end{figure}

To illustrate how friction creates final state sensitivity, we consider
a classical particle moving along a straight line in the presence
of a symmetric double-well potential. The system is described by the
Hamiltonian function 
\begin{equation}
H(p,q)=\frac{p^{2}}{2m}+W(q)\,,\quad p,q\in\mathbb{R\,},\label{eq: Hamiltonian}
\end{equation}
where $q$ and $p$ denote position and the momentum of the particle,
respectively. The minima of the potential $W(q)$ are located at $q=q_{\pm}$
, separated by a barrier of height $W_{0}\equiv W(0)$ which defines
the \emph{critical} energy, $E_{c}\equiv W_{0}$. The phase-space
diagram of the system is shown in Fig. \ref{fig: double well flow (general)}, displaying 
the familiar types of trajectories. The minima $L$ and $R$ of the
potential are \emph{stable fixed points} each surrounded by \emph{periodic
orbits} with energies $E$ not exceeding the critical value, $0<E<E_{c}\equiv W_{0}$.
For $E=E_{c}$, the particle may rest at the \emph{unstable fixed
point} at $q=0$, or travel on one of the two \emph{separatrices}
connected to it. The trajectories with energy above the critical value,
$E>E_{c}$ are \emph{periodic}, encircling both minima on a single
round trip. With a single degree of freedom and the energy $H(p,q)$
as a conserved quantity, the system is \emph{integrable}, leading
to the global foliation of its phase space into one-dimensional tori. 

Adding friction will modify all trajectories except when then particle
initially rests at one of the three fixed points. If located on a
separatrix or on any periodic trajectory with energy less than $E_{c}$,
dissipation will cause the particle to ``spiral'' into either the
left or the right minimum of the potential $W(q)$, depending on its
original position relative to the origin, $q=0$. The particle cannot
escape from a well once it has been trapped, and the fixed points
$L$ and $R$ turn into \emph{attractors}. 

The destiny of a particle with initial energy $E>E_{c},$ however,
it is not immediately obvious since it may end up in either well.
Friction will inevitably ``draw'' the particle towards the location
of the separatrices of the unperturbed system. At some time, the energy
of the particle will drop below the critical value $E_{c}$. The position
of the particle relative to the origin at the time of the drop will
determine whether it becomes trapped in the left or in the right well.

For simplicity, let us assume that friction acts at discrete times
only, repeatedly reducing the momentum of the particle by a constant
factor. Suppose that for initial conditions $\mathbf{z}=(p,q)^{T}$,
the particle will -- after a possibly long time -- settle in the \emph{right}
well as illustrated in Fig. \ref{fig: double well flow (general)}.
Intuitively, a slightly smaller initial momentum (see $\mathbf{z}^{\prime}$
in the figure) could cause the particle to negotiate the barrier one
less time and to settle in the \emph{left} well instead. The slight
change in the initial condition has thus altered the long-term behaviour
of the system. Therefore, the finally state of a particle may become
unpredictable from a \emph{practical} point of view, i.e. whenever
its initial conditions are known to lie within a small but finite
volume of phase space only. 

More formally, the non-Hamiltonian equations of motion map an initial
state $\mathbf{z}(t_{0})$ to a new value $\mathbf{z}(t)$ at time
$t$,
\begin{equation}
\mathbf{z}(t_{0})\mapsto\mathbf{z}(t),\quad\mathbf{z}\equiv\begin{pmatrix}p\\
q
\end{pmatrix}\,,\label{eq: general dynamics}
\end{equation}
leading to a \emph{decrease} of the energy defined in (\ref{eq: Hamiltonian}):
$E_{0}\to E<E_{0}$. To ascertain whether a particle with initial
state $\mathbf{z}(t_{0})$ ends up near $L$ or near $R$, one needs
to determine the earliest time $t$ such that its energy $E$ falls
below the critical value,
\begin{equation}
E<E_{c}\,.\label{eq: energy condition}
\end{equation}
Repeating this calculation for \emph{all} initial conditions will
divide the phase space into two disjoint sets known as \emph{basins
of attraction }which encode whether the particle ends up in well ${\cal L}$
or ${\cal R}$. Let us investigate the structure of their boundaries
for two models of friction, using a particularly simple double-well
potential.

\section{Piece-wise constant double-well potential with friction\label{sec:Model-systems}}

The double-well potential considered here is based on a ``particle
in a box'' defined by two infinitely high potential walls at $q=\pm\ell$
which restrict motion to a line segment of length $2\ell$. The particle
bounces off the walls elastically resulting in an instantaneous reversal
of its momentum: $p\to-p$; its position $q=\pm\ell$ remains unchanged
when hitting a wall. A piece-wise constant potential,

\begin{equation}
W(q)=\begin{cases}
W_{0}\,, & |q|<\varepsilon\,,\\
0\,, & \varepsilon<|q|<\ell\,,
\end{cases}\label{eq:flat double well}
\end{equation}
models the smooth double-well. For simplicity, we take an arbitrarily
thin potential barrier, corresponding to $\varepsilon\to0$. The only
impact of this ``infinitesimal'' barrier is to confine the particle
in a well once its energy drops below the critical value $W_{0}$,
thus creating the wells ${\cal L}$ and ${\cal R}$. Two continuous
sets of potential minima exist because the bottom of the potential
is flat. 

A widespread method to investigate \emph{non-integrable} systems is
to start with an integrable system and add a perturbation, be it a
time-independent potential term as in the KAM theorem or a deterministic
time-dependent force \cite{chirikov79}. To support our claim that
friction generically causes final state sensitivity, we will model
it in two different ways which are inspired by these approaches. In
the first case, the elastic collisions of the particle with the boundary
walls are made inelastic (cf. Sec. \ref{sub:Inelastic-collisions})
while an impulsive friction force is applied periodically in the second
case (cf. Sec. \ref{sub:Periodic-damping}). The first model depends
on a single parameter only, the \emph{coefficient of restitution.}
The second model depends on two parameters, the \emph{frequency} and
the \emph{strength} of the dissipative ``kick.''

\subsection{Inelastic collisions\label{sub:Inelastic-collisions}}

The motion of the particle in the piece-wise constant double well
(\ref{eq:flat double well}) consists of free motion between the walls
interspersed with momentum-reversing elastic collision at the walls.
The dynamics changes fundamentally upon replacing the elastic collisions
at the walls by inelastic ones, characterized by a \emph{coefficient
of restitution}, $r\in(0,1)$:
\begin{equation}
\mathbf{z}\mapsto\mathbf{z}^{\prime}=\mathbf{R\cdot}\mathbf{z}\text{\,,\quad\ensuremath{\mathbf{R}\equiv\left(\begin{array}{cc}
 -r  &  0\\
0  &  1 
\end{array}\right)}\qquad for}\quad q=\pm\ell\,.\label{eq: restitution}
\end{equation}
This minor change turns the conservative system into a dissipative
one and -- as we will see -- is sufficient to create an embryonic
form of final-state sensitivity. 

The long-term dynamics of the particle does not depend on its initial
\emph{position}: all particles with fixed momentum $p_{0}$ but arbitrary
position $q_{0}\in(-\ell,\ell)$ will experience the same amount of
friction, only to end up in same well. Thus, let us assume that the
particle starts out with positive initial momentum $p_{0}>p_{c}$,
beings located at $q_{0}=\ell_{-}$, i.e. just to left of the right
wall. Then, the initial state $\mathbf{z}_{0}=(p_{0},\ell_{-})$ at
time $t_{0}$ evolves according to 
\begin{equation}
\mathbf{z}(t_{n})=\mathbf{R}^{n}\cdot\mathbf{z}_{0}=\begin{pmatrix}(-r)^{n}p_{0}\\
\ell_{-}
\end{pmatrix}\,,\qquad n\in\mathbb{N}_{0}\,,\label{eq: restitution evolution}
\end{equation}
with the times $t_{n}$ being defined by particle returning to its
initial position $\ell_{-}$. Monitoring the value of its momentum
at the walls is sufficient to determine the well which will trap the
particle. The particle will be trapped in well ${\cal R}$, for example,
if its last collision at the right wall makes its energy drop below
the critical value $E=E_{c}$ due to $p\mapsto(-r)p$. 

\begin{figure}[h]
\begin{centering}
\includegraphics[bb=-140bp 0bp 1200bp 500bp,scale=0.5]{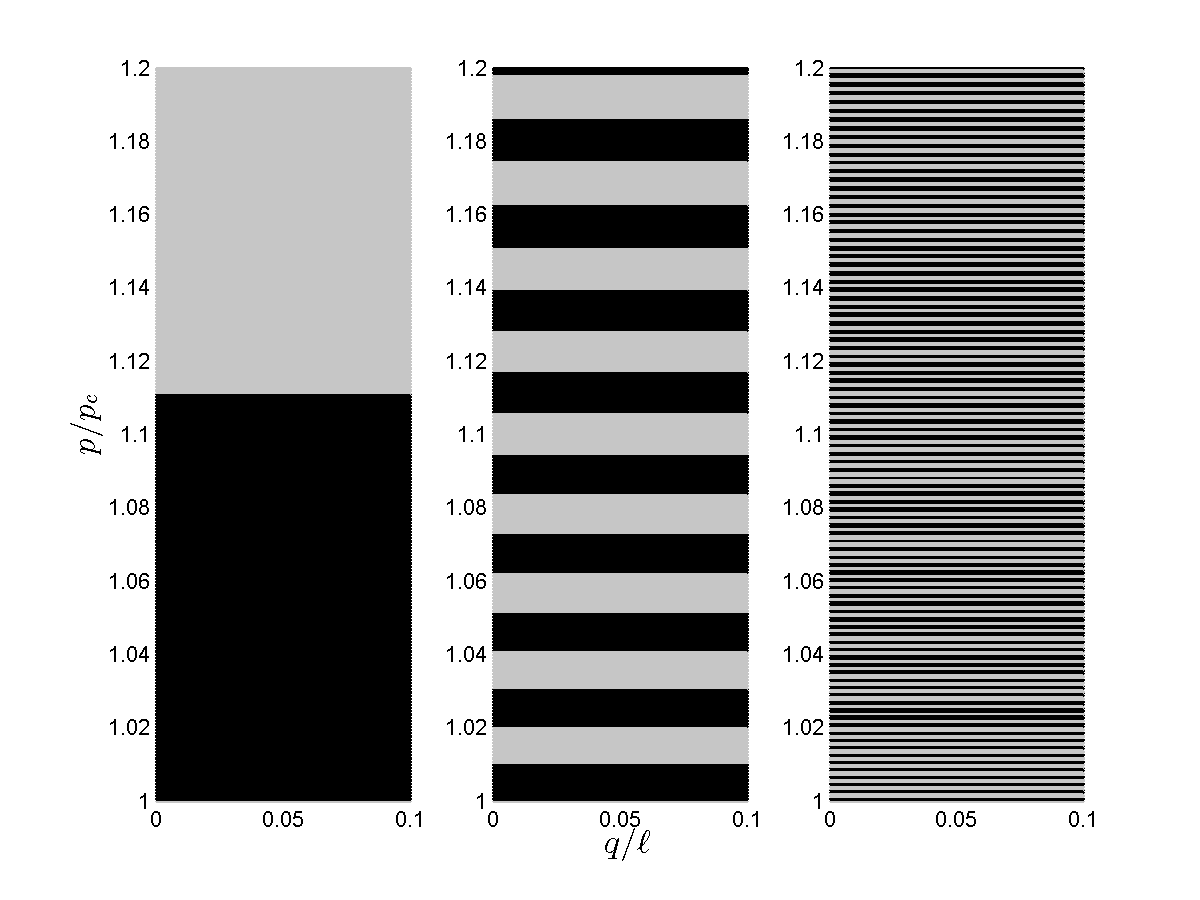} \protect\caption{Basins of attraction for the phase-space region $(1\leq p/p_{c}\leq1.2$,
$0\leq q/\ell\leq0.1$) of a particle of unit mass in a double-well,
with friction arising from inelastic collisions at the boundary walls:
initial conditions located in dark (light) regions will end up in
the right (left) well of the potential. The vertical bars correspond
to different values of the coefficient of restitution: $r=0.9$, $r=0.99$,
$r=0.999$ (left to right). Smaller values of friction lead to increasing
``complexity'' of the basin boundaries, in the sense of producing
narrower bands. \label{fig: basins of attraction (inelastic wall model)}}
\par\end{centering}
\end{figure}

For positive initial momentum $p_{0}$ the particle will hit the right
wall first. The well to finally trap the particle is determined by
the number of collisions $n(E_{0})$ before it drops below $E_{c}$.
Denoting the energy of the particle after $n$ collisions by $E_{n}$,
we need to find the number $n(E_{0})$ such that the energy of the
particle falls below $E_{c}$ for the first time,
\begin{equation}
n(E_{0})=\min_{n\in\mathbf{N}}\,\left\{ E_{n}<E_{c}\right\} \,,\quad\mbox{with }E_{0}>E_{c}\,.\label{eq: energy constraint (inelastic)}
\end{equation}
Using Eq. (\ref{eq: restitution evolution}) the number $n(E_{0})$
is easily found to be 
\begin{equation}
n(E_{0})=\left\lceil \frac{1}{2}\frac{\ln\left(E_{c}/E_{0}\right)}{\ln r}\right\rceil =\left\lceil \frac{\ln\left(p_{c}/p_{0}\right)}{\ln r}\right\rceil \,,\label{eq: LR condition}
\end{equation}
where the initial moment defines the initial energy, $E_{0}=p_{0}^{2}/2m$,
and $\left\lceil x\right\rceil $ is the ceiling function extracting
the smallest integer greater or equal to the number $x$. If $n$
is odd (even), a particle with positive momentum $p_{0}$ will end
up in the well on the right (left). The basins of attraction for the
wells ${\cal L}$ and ${\cal R}$ are given by alternating horizontal
bands in phase space shown in Fig. \ref{fig: basins of attraction (inelastic wall model)}.
The widths of the bands decrease with decreasing friction (and they
increase with energy $E$ which the figure does not show due to the
limited momentum range).
If the initial conditions $(p_{0},q_{0})$ of a particle are known
\emph{exactly}, then the deterministic dynamics leads to a unique
and well-defined final state which can be predicted with certainty.
However, limited precision of the initial conditions may results in
a genuine indeterminacy of the final state. Assume that the initial
state of the particle is only known to lie inside a rectangle with
sides $\Delta q>0$ and $\Delta p>0$, centered about the point $\mathbf{z}_{0}$.
Trajectories with initial momenta $p_{0}$ and $p_{0}^{\prime}\equiv rp_{0}$
are bound to end up in different wells. Thus, if the inaccuracy in
momentum exceeds this value,
\begin{equation}
\Delta p>p_{0}-p_{0}^{\prime}\,,
\end{equation}
the uncertainty rectangle will cut across \emph{at least two} adjacent
basins of attraction. In other words, given the initial momentum $p_{0}$
and any finite uncertainty $\Delta p$ about it, the prediction of
the final state becomes impossible for a coefficient of restitution
in the interval 
\begin{equation}
1-\frac{\Delta p}{p_{0}}<r<1\,,\label{eq: interval for r}
\end{equation}
since the rectangle with sides $\Delta q$ and $\Delta p$ will contain
trajectories destined for the wells ${\cal L}$ and ${\cal R}$. We
conclude that sufficiently weak inelasticity prevents the reliable
prediction of the final state. In this well-defined sense, adding
friction to an integrable system provides a mechanism which prevents
accurate long-term predictions.

\subsection{Periodic damping\label{sub:Periodic-damping}}

Now we turn to a model where friction is caused by a periodic, dissipative
force which acts during a short time interval only. It will be convenient
to consider the limit of an \emph{instantaneous} action which multiplies
the momentum of the particle by a constant factor $\gamma\in(0,1)$
at times $T_{k}=kT$, with $k\in\mathbb{N}$, and a free parameter
$T$. This approach is analogous to periodically kicking a system
with a deterministic force which, for a particle moving freely on
a ring known as a ``rotor,'' produces \emph{deterministically chaotic}
motion \cite{arnol'd89}. Since our model depends on \emph{two }independent
parameters, $\gamma$ and $T$, we expect more complicated basins
of attraction compared to the model with inelastic reflections.

To construct the basins of attraction of the wells ${\cal L}$ and
${\cal R}$, we need to determine when, for arbitrary initial conditions
$(p_{0},q_{0})^{T}$, the energy of the particle falls below the critical
value for the first time We then record whether, at that moment of
time, it is located to the left or to the right of the origin, i.e.
within ${\cal L}$ or ${\cal R}$. For simplicity, the particle is
assumed to begin its journey at time $t=0^{+}$, i.e. just after $t=0$,
with \emph{positive} momentum $p_{0}>p_{c}$ and arbitrary initial
position $q_{0}\in(-\ell,\ell$). 

The particle moves freely during intervals of length $T$, with perfectly
elastic collisions occurring at the boundary walls which only change
the sign of its momentum. An expression for its time evolution in
closed form can be found if we ``unfold'' the trajectory by imagining
identical copies of the double-well to be arranged along the position
axis. Instead of being reflected at the right wall located at $q=\ell$,
the particle enters the next double well, which occupies the range
$(\ell,3\ell$), and continues to move to the right, etc. In this
setting, the momentum does not change its sign when the particle moves
from one double well to the adjacent one. The sign of its momentum
in the original double well is negative (or positive) if the particle
has reached the $s^{th}$ copy of the double well, with $s\in\mathbb{N}$
being odd (or even).

To determine the dynamics of the system over one period of length
$T$, we combine the free motion with the dissipative kicks: 
\begin{enumerate}
\item during the motion of the particle from $t=0^{+}$ to just before the
first kick at time $t=T$, its phase-space coordinates are given by
\begin{equation}
\mathbf{z}(t)=\begin{pmatrix}p_{0}\\
q_{0}+p_{0}t/m
\end{pmatrix}\equiv\mathbf{F}(t)\text{\ensuremath{\cdot}}\mathbf{z}_{0}\,,\qquad\mathbf{F}(t)=\begin{pmatrix}1 & 0\\
t/m & 1
\end{pmatrix}\,,\qquad t\in(0^{+},T^{-})\,,\label{eq: free motion}
\end{equation}
where $q\in(0,\infty)$ due to the unfolding; 
\item the dissipative kick at time $T$ reduces the momentum of the particle
by the factor $\gamma\in(0,1)$,
\begin{equation}
\mathbf{z}(T^{+})=\begin{pmatrix}\gamma p(T^{-})\\
q(T^{-})
\end{pmatrix}\mathbf{\equiv D}\cdot\mathbf{z}(T^{-})\,,\qquad\mathbf{D}=\left(\begin{array}{cc}
\gamma\  & 0\\
0 & 1
\end{array}\right)\,,\qquad t\in(T^{-},T^{+}).\label{eq: first kick}
\end{equation}

\end{enumerate}
To obtain the \emph{actual} position and momentum of the particle
inside the box at time $t$, we map (or ``fold back'') the expression
$\mathbf{F}(t)\cdot\mathbf{z}$ to the interval $q\in(-\ell,\ell)$,
by writing

\begin{equation}
\mathbf{z}(t)=\begin{pmatrix}(-)^{s(t)}p_{0}\\
\left[\left(q_{0}+p_{0}t/m\right)\!\!\!\!\!\!\mod2\ell\right]-\ell
\end{pmatrix}\,,\qquad t\in(0^{+},T^{+})\,,\label{eq: actual coordinates after one period}
\end{equation}
where the value of the integer $s(t)$ is determined by writing $q+pt/m=\overline{q}(t)+2\ell s$,
with $\overline{q}(t)\in(-\ell,+\ell)$. The momentum $p$ changes
sign whenever the ``unfolded'' coordinate passes through the values
$\ell,3\ell,5\ell,\ldots$

The time evolution of the initial state $\mathbf{z}_{0}$ from time
$t=0^{+}$ to $t=T_{k}^{+}\equiv(kT)^{+}$, i.e. just after the kick
with label $k$, follows from concatenating Eqs. (\ref{eq: free motion})
and (\ref{eq: first kick}) $k$ times,
\begin{equation}
\mathbf{z}(T_{k}^{+})=\left(\mathbf{D}\cdot\mathbf{F}(T^{-})\right)^{k}\cdot\mathbf{z}_{0}\equiv\begin{pmatrix}\gamma & 0\\
\gamma T/m & 1
\end{pmatrix}^{k}\cdot\mathbf{z}_{0}=\begin{pmatrix}\gamma^{k} & 0\\
\sigma_{k}(\gamma)T/m & 1
\end{pmatrix}\cdot\mathbf{z}_{0}\,,\label{eq: periodic damping time evolution}
\end{equation}
where
\begin{equation}
\sigma_{k}(\gamma)=\frac{1-\gamma^{k}}{1-\gamma}\,,\qquad k\in\mathbb{N}\,.\label{eq: geometric series for gamma}
\end{equation}
In analogy to Eq. (\ref{eq: actual coordinates after one period}),
the ``true'' coordinates of the particle inside the box are obtained
as
\begin{equation}
\mathbf{z}(T_{k}^{+})=\begin{pmatrix}(-)^{s(t)}\gamma^{k}p_{0}\\
\left[\left(q_{0}+\sigma_{k}(\gamma)p_{0}T/m\right)\mod2\ell\right]-\ell
\end{pmatrix}\,,\label{eq: actual coordinates after k periods}
\end{equation}
assuming that, after $k$ kicks, the energy $E_{k}=p_{k}^{2}/2m$
of the particle has not yet dropped below the critical value $E_{c}$. 

We are now in the position to determine which initial conditions $\mathbf{z}_{0}$
will send the particle to the left and the right well, respectively.
Using Eq. (\ref{eq: actual coordinates after k periods}), we first
determine the smallest value of $k$ which reduces the energy of the
particle below the critical value, $E_{k}<E_{c}$, or
\begin{equation}
k_{c}=\left\lceil \frac{1}{2}\frac{\ln(E_{c}/E_{0})}{\ln\gamma}\right\rceil =\left\lceil \frac{\ln(p_{c}/p_{0})}{\ln\gamma}\right\rceil ,\qquad k_{c}\in\mathbb{N}\,,\label{eq: critical k (periodic damping)}
\end{equation}
assuming, of course, that $p_{0}>p_{c}$. This relation structurally
resembles the result (\ref{eq: LR condition}), with the number $k_{c}$
of dissipative kicks playing the role of the number of inelastic collisions
$n_{c}.$ The sign of the position coordinate after $k_{c}$ kicks,
$q(T_{k_{c}}^{+})$, follows from Eq. (\ref{eq: actual coordinates after k periods})
and determines whether the particle is trapped in ${\cal L}$ or ${\cal R}$.
The explicit dependence of $\mathbf{z}(T_{k}^{+})$ on the initial
\emph{position} implies that changes in $q_{0}$ may also produce
different final states, in contrast to the model studied in Sec. \ref{sub:Inelastic-collisions}.

\begin{figure}[h]
\centering{}\includegraphics[bb=-130bp 0bp 1200bp 500bp,scale=0.5]{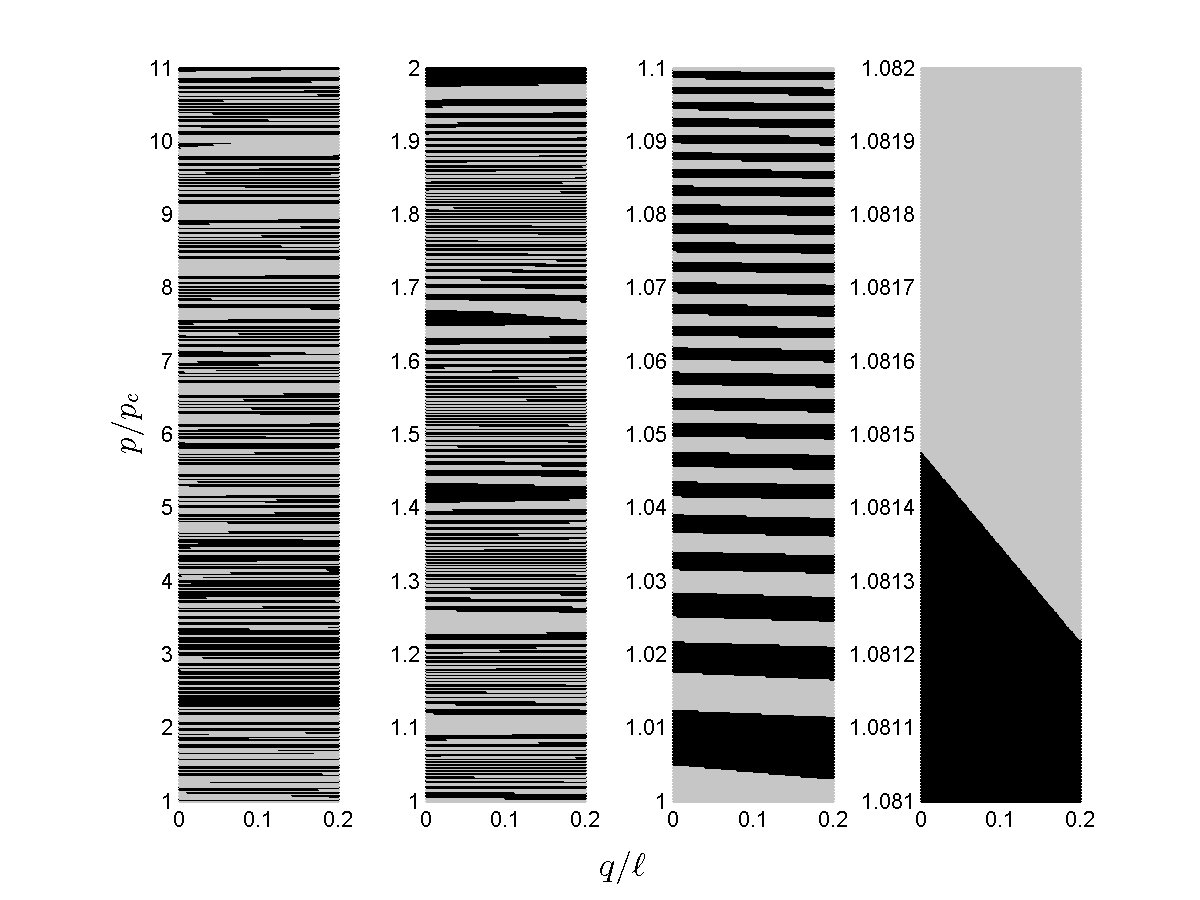}\protect\caption{Basins of attraction for the phase-space region $(1\leq p/p_{c}\leq11$,
$0\leq q/\ell\leq0.2$) of a particle with unit mass in a double-well,
with friction arising from \emph{periodic dissipative kicks }at times\emph{
$kT$, }$k\in\mathbb{N}$, with $\ell=1$,\emph{ $T=100s$} and $\gamma=0.99$:
initial conditions located in dark (light) regions will end up in
the right (left) well of the potential. Each of the three vertical
bars on the right magnifies a horizontal strip of the bar to its left
by a factor of ten. The basin boundaries clearly exhibit both a momentum
\emph{and} a position dependence. \label{fig: magnified basins (kicked)}
Each bar results from iterating $501\times501$ regularly spaced initial
conditions inside the area shown. }
\end{figure}

Fig. \ref{fig: magnified basins (kicked)}, which has been generated
numerically on the basis of Eq. (\ref{eq: actual coordinates after k periods}),
illustrates these conclusions. The first vertical bar visualizes the
basins of attraction associated with the wells $\mathcal{L}$ and
$\mathcal{R}$, respectively. The expected dependence on both initial
momentum and position becomes clearly visible in the magnifications
which also reveal that the boundaries of the apparently irregular
basins of attractions are not fractal. 

The boundaries of the basins can be found directly from Eq. (\ref{eq: actual coordinates after k periods}):
all initial conditions $(p_{0},q_{0})$ mapped to a fixed value of
position at time $kT$ are located on lines of the form
\begin{equation}
p_{0}(q_{0})=-\frac{m}{\sigma_{k}(\gamma)T}q_{0}+\mbox{const}\simeq-\frac{1}{k}\frac{m}{T}q_{0}+\mbox{const}\,,\qquad\gamma\apprle1\,,
\end{equation}
using $\sigma_{k}(\gamma)\simeq(1/k)+{\cal O}(1-\gamma)$, which holds
for weak damping, i.e. for $\gamma$ approaching the value one from
below. Consequently, the boundaries of the basins of attraction are
\emph{straight lines }in phase space just as for the model with inelastic
reflections off the wall. The lines are no longer horizontal but their
slope approaches the value zero if a large number of kicks is required
for the particle to settle in a well. 

Assume once again that the initial conditions of the particle can
be prepared with \emph{finite precision} only, i.e. they lie inside
a phase-space rectangle with area $\Delta q\Delta p>0$ and center
$\mathbf{z}_{0}$. For any finite imprecision one can always find
a damping strength $\gamma$ such that at least one basin boundary
crosses the rectangle; this is sufficient to prevent the prediction
of the well to finally trap the particle. For large initial momenta
$p_{0}$, the reasoning behind the derivation of the inequality (\ref{eq: interval for r})
also applies here since the strips constituting the basins of attraction
will, typically, have almost horizontal boundaries. Thus, for any
initial conditions $(p_{0},q_{0})$ and finite uncertainties, damping
strengths within the interval
\begin{equation}
1-\frac{\Delta p}{p_{0}}<\gamma<1\label{eq: interval for gamma}
\end{equation}
correspond to a situation with an unpredictable final state. Occasionally,
the uncertainty rectangle with sides $\Delta q$ and $\Delta p$ may
cover an area where a slight change in initial \emph{position} causes
the particle to reach different wells, which only increases the final
state sensitivity.

\section{Summary and conclusions\label{sec:Summary}}

We have shown that adding friction to an integrable one-dimensional
double-potential well causes its dynamics to exhibit a rudimentary
form of final-state sensitivity. For simplicity, the well has been
modeled as a ``box'' divided into two regions by a thin wall. A
particle has been subjected to two types of dissipative forces which,
by reducing its initial energy, cause the particle to necessarily
settle in one the wells after a finite, possibly long time. The main
result of our study is that adding friction to an integrable system
produces basins of attraction with finely structured boundaries. 

If the particle collides \emph{inelastically }with the confining walls
the resulting basins foliate the phase space of the system into horizontal
layers of variable width which get narrower for decreasing friction.
Any ball of initial conditions which extends beyond more than one
band prevents us from predicting with certainty the well in which
the particle will finally settle. \emph{Periodic dissipative kicks}
create basins of attraction with slightly more intricate boundaries,
due to their additional position dependence. Since the particle must
settle in a well after \emph{finite} time the observed structures
cannot be fractal. In practice, however, it is crucial whether the
initial conditions can be specified with sufficient accuracy to avoid
a spread across basins which send the particle to different final
states.

These model systems demonstrate that adding friction to an integrable
system with multiple stable equilibria can have a fundamental impact
on long-term predictability. The motion is not ``deterministically
random'' which would require fractal phase-space structures. However,
if the accuracy of the initial conditions falls below a specific threshold,
the final state of the system cannot be predicted reliably. Experimentally,
the precision required for a reliable long-term prediction may well
be out of reach.

We expect our conclusions to be structurally stable in the sense that
they should not depend on the model of friction used. Any dissipative
mechanism will, firstly, contract all initial conditions into a small
phase-space region which is energetically just above the barrier of
the double well; secondly, the energy of the particle will drop below
$E_{c}$ in a way which depends sensitively on the initial conditions.
Continuous Stokes friction, for example, is thus likely to generate
similar basins of attraction.

To systematically study the creation of basins of attraction with
intricate boundaries in more general, smooth potential wells, we suggest
to exploit the existence of action-angle variables $(I,\varphi)$
in integrable systems. The energy $E$ represents a convenient starting
point to study the impact of friction forces, when expressed as a
function of the action $I$, 
\begin{equation}
I(E)=\oint p(q,E)dq\,.\label{eq: general action}
\end{equation}
Suitable perturbations are easily added to the new form of the Hamiltonian,
$H=H(I)$, once position $q$ and momentum $p$ have been mapped to
$(I,\varphi$) by means of a canonical transformation. 

Finally, we highlight a natural application of the effective unpredictability
of a final state due to friction, given sufficiently imprecise initial
conditions. It arises upon introducing a larger number of identical
potential wells arranged on a ring, (37 or 38, say), mimicking a one-dimensional
roulette wheel. Including periodic dissipative kicks provides a surprisingly
simple explanation why a finite spread in initial momenta and positions
is sufficient to generate random outcomes, the working hypothesis
underlying any gambling. This approach should be contrasted with models
of an actual roulette wheel where unpredictable trajectories arise
through a multitude of effects: motion in a bent annulus-shaped region
embedded into three dimensions, the presence of gravity, rolling resistance
and a (presumably) non-integrable time-dependent potential.

\subsection*{Acknowledgments}

JO is grateful for financial support through a ``Summer-2014 Publication
Studentship'', provided by the Department of Mathematics at the University
of York, UK.

\end{document}